%% file: main.tex
\documentclass[conference,letterpaper]{IEEEtran}
\usepackage[letterpaper, left=0.62in, right=0.62in, bottom=1.2in, top=0.8in]{geometry}

\usepackage{romannum}
\usepackage{graphicx}
\usepackage{multirow}
\usepackage{amssymb,bm}
\usepackage{bbm}
\usepackage{algpseudocode}
\usepackage{algorithm}
\usepackage{array}
\usepackage{color}
\usepackage[utf8]{inputenc}
\usepackage[T1]{fontenc}
\usepackage{url}
\usepackage{ifthen}
\usepackage{cite}
\usepackage{mathtools}
\usepackage[english]{babel}
\usepackage{amsthm}
\usepackage{thmtools}
\usepackage{enumerate}
\usepackage{caption}
\usepackage{subcaption}
\usepackage[acronym]{glossaries}

\usepackage{titlesec}
\titlespacing*{\section}{0pt}{0.1\baselineskip}{0.2\baselineskip}

\declaretheoremstyle[%
  spaceabove=6pt,%
  spacebelow=6pt,%
  headfont=\normalfont\itshape,%
  bodyfont=\normalfont,%
  postheadspace=1em,%
  qed=\qedsymbol%
]{mystyle}

\declaretheoremstyle[%
  spaceabove=6pt,%
  spacebelow=6pt,%
  headfont=\bfseries,%
  bodyfont=\normalfont,%
  postheadspace=1em,%
]{mystyle_1}

\ifCLASSINFOpdf
\else

\fi

\begin{document}
\pagenumbering{arabic}
\title{A Lossless Compression Technique for the Downlink Control Information Message}

\author{
  \IEEEauthorblockN{Bryan~Liu,~\IEEEmembership{Member, ~IEEE,}
  Alvaro Valcarce,~\IEEEmembership{Senior Member, ~IEEE,}
  and
  K. Pavan Srinath ~\IEEEmembership{Member, ~IEEE} \\}
  \IEEEauthorblockA{Nokia Bell Labs, Massy, France \\
                    Email: \{bryan.liu, alvaro.valcarce\textunderscore rial, pavan.koteshwar\textunderscore srinath\}@nokia-bell-labs.com}

}

\maketitle

\input{acronyms}

\begin{abstract}
Improving the reliability and spectral efficiency of wireless systems is a key goal in wireless systems.
However, most efforts have been devoted to improving data channel capacity, whereas control-plane capacity bottlenecks are often neglected.
In this paper, we propose a means of improving the control-plane capacity and reliability by shrinking the bit size of a key signaling message - the 5G \gls{dci}. In particular, a transformer model is studied as a probability distribution estimator for Arithmetic coding to achieve lossless compression.
Feature engineering, neural model design, and training technique are comprehensively discussed in this paper.
Both temporal and spatial correlations among \gls{dci} messages are explored by the transformer model to achieve reasonable lossless compression performance. Numerical results show that the proposed method achieves $~21.7\%$ higher compression ratio than Huffman coding in \gls{dci} compression for a single-cell scheduling scenario.

\end{abstract}
\IEEEpeerreviewmaketitle

\vspace{6pt}
\section{Introduction}
In wireless systems, the control plane suffers capacity bottlenecks when a large number of devices with low user-plane traffic are in the network.
In particular, 6G networks will need to handle a massive number of devices and it is expected that more device types will connect to the 6G network than to the current 5G system \cite{sixG}. 
The conventional solution could be to enhance the reliability of the control channel through a stronger detection algorithm or by implementing a stronger channel code \cite{blind_detection, polar_code_1}, so that the number of re-transmissions is reduced. 
However, the number of resources is limited, and enhancing the physical framework or algorithms does not solve the root problem of packing the control message and transmitting it through the wireless medium with finite resources. 
To precisely manage a large number of devices, the control channel must optimize resource usage. 
Accurately transmitting the source downlink/uplink control information to the receiver side with a low overhead delivers a larger capacity. 
A shorter control message would require less radio resources so that more devices can be served in a limited time. Further, a smaller payload can profit from additional error-correcting bits, thus improving the reliability of the channel. 
Pursuing that line of thought, reducing the control message length becomes an efficient way of improving the capacity of a system and reducing the overall transmission latency.
In 5G New Radio (NR), 
\gls{dci} messages are independently generated between consecutive subframes. 
However, in real systems, the behavior of devices usually follows recognizable patterns. 
This possibly creates correlations amongst the \gls{dci} messages which can be exploited by data-based techniques to improve the efficiency of \gls{dci} messaging.

State-of-the-art methods for reducing the length of a message can be divided into lossless and lossy categories.
Lossy compression introduces recovery loss, while lossless compression guarantees that the encoded message can be decoded into the original message error-free. Since accurately decoding a control message is necessary for maintaining a reliable and stable wireless system, lossless compression is the preferred choice for reducing the length of a \gls{dci} message.
Recent lossless compression techniques \cite{deepzip} employ \gls{ml} to learn the underlining distribution of the source data and achieve a better compression ratio compared to traditional ``look-up'' table-based methods such as \gls{hc} \cite{huffmanCoding} and Lempel-Ziv-Welch \cite{lempel_ziv_welch}. 
However, the application of these methods to control-plane signaling compression in wireless systems does not exist in the literature to the best of the authors' knowledge.

This paper studies the problem of reducing the \gls{dci} message bit-size in wireless communication systems. We propose transformer-based encoders and decoders for the lossless compression of \gls{dci} messages with the assistance of \gls{ac}. The spatio-temporal correlation amongst \gls{dci} messages is exploited by the transformer model. 
Specifically, the feature embedding, the \gls{nn} architecture, and the training techniques are presented in this paper. Besides demonstrating reasonable compression performance and outperforming the baselines while remaining 5GNR-compliant, the channel decoding performance of Polar codes under different compression techniques is evaluated to demonstrate the potential improvement in the reliability of Physical Downlink Control Channel (PDCCH) with DCI compression.
Numerical results show that a transformer-based DCI compression technique can achieve approximately $21.7\%$ improvement in compression ratio compared to HC for a simple single-cell wireless network.

\section{Preliminaries}
\subsection{Downlink control information}
\gls{dci} carries the control-plane signaling from the \gls{bs} to the \gls{ue} through a \gls{pdcch}.
It is an essential message needed by the \gls{ue} to successfully decode the data packets.

Multiple \gls{dci} formats exist, and each is used depending on the control commands that the \gls{bs} needs to convey to the \gls{ue} \cite{dci_3gpp} at a given point in time. The \gls{dci} message can contain information fields such as resource allocation, modulation and coding scheme, and power control commands. At the receiver side, the \gls{ue} blindly searches the PDCCH search-space-sets and decodes PDCCH candidates until a \gls{crc} has passed. It then reads the \gls{dci} and follows the \gls{bts} commands from the decoded \gls{pdcch}.
Since the channel and traffic models for a UE might follow a certain trend, the \gls{dci} message to be sent at the current \gls{tti} is likely to be correlated to past \gls{dci} messages. Besides being correlated in time, the fields within one \gls{dci} message may also be correlated. 
As a result, ``repeated'' information is transmitted which allows a good lossless compression algorithm to reduce the length of \gls{dci} messages and consequently provide control information to a massive number of \gls{ue} under time constraints.

\vspace{-2mm}
\subsection{Deep-learning based lossless compression}
\vspace{-2mm}
The framework proposed in \cite{deepzip} employs a \gls{rnn} to explore the correlation between consecutive symbols to compress a sequence of symbols and it can be summarized by the following two steps:
\begin{itemize}
	\item \textbf{Probability estimation}: Conditioned on the information from previously encoded (decoded) symbols, an \gls{rnn} model estimates the probability distribution of the current symbol to be encoded (decoded).
	\item \textbf{Arithmetic coding}: Taking the probability distribution returned from the \gls{rnn} as an input, \gls{ac} encodes (decodes) the symbol into binary bits in a way such that frequently used symbols are encoded with fewer bits.
\end{itemize}

Under an AC framework, better distribution estimates of the DCI fields conditioned on the input features will lead to better compression ratios. 
To train the \gls{nn}, a categorical cross-entropy loss between the actual label of each symbol and the \gls{rnn} estimate is used. \glspl{rnn} capture temporal correlations through a hidden state vector but struggle with long-range dependencies due to their recurrent structure.
Implementing \glspl{rnn} in wireless communication systems for control message compression is challenging due to the correlations across multiple \glspl{tti}.

Recent studies demonstrate that transformer models \cite{transformer} effectively address long-range dependencies and can be adapted for lossless compression in \cite{deepzip} to create a transformer-based algorithm for this purpose \cite{transformer_deepzip}.
However, the application of transformers specifically for \gls{dci} compression in wireless communications is not straightforward and requires addressing aspects such as field embedding, compression steps, and feature selection. The subsequent sections will detail one such transformer-based \gls{dci} compression method.

\section{Transformer-based lossless compression for DCI message}
Let $\mathbf{X} = \{\mathbf{x}_1, \mathbf{x}_2,...,\mathbf{x}_T \}$ represent the set of \gls{dci} messages collected from the first \gls{tti} to the $T$-th \gls{tti}. Each \gls{dci} message $\mathbf{x}_t \triangleq \{x_{t, 1}, x_{t, 2}, ..., x_{t, N}\}$ contains $N$ bits, where $x_{t, i} \in \mathbb{F}_2$. Furthermore, a \gls{dci} message is structured into $D$ distinct fields. For a given message in the $t$-th TTI, the $k$-th field is denoted by $\mathbf{d}_{t,k}$ so that we also have $\mathbf{x}_t = \{\mathbf{d}_{t,1}, \mathbf{d}_{t,2}, ..., \mathbf{d}_{t,D}\}$. Each field $\mathbf{d}_{t,k}$ consists of $M_k$ bits, and hence, we have $\sum_{k=1}^D M_j = N$. For the convenience of concept interpretation, we define the following 2 terms of correlation that could exist within \gls{dci} control bits.
\begin{itemize}
\vspace{-0.5mm}
	\item \textbf{Temporal correlation}: The absolute of the Pearson correlation coefficient, $|\rho(x_{t,i}, x_{t-\hat{t}, j})|$, might be greater than 0 for $\hat{t} < t$ with $i, j \in \{1,2,...,N\}$. Temporal correlation refers to the possible correlation between the current and the previous control bits in the \gls{dci} control messages.
	\item \textbf{Spatial correlation}: The absolute of Pearson correlation coefficient, $|\rho(x_{t,i}, x_{t, j})|$, might be greater  than 0 for any $i\neq j$. Spatial correlation refers to the possible correlation amongst the control bits in the current \gls{tti}.
 \vspace{-1mm}
\end{itemize}

\vspace{-4pt}
\subsection{Features construction}
\vspace{-1mm}
The main motivation for employing a \gls{nn} predictor for \gls{ac} is to learn the underlying correlation between symbols to accurately estimate the probability distribution of each symbol to be compressed. If the \gls{dci} message is compressed following the method proposed in \cite{deepzip}, a direct way of constructing the input features for the \gls{nn} predictor would be a concatenated sequence $\mathbf{u}_{t,i} = \bigl[\mathbf{x}_{t-L}, \mathbf{x}_{t-L+1},..., \mathbf{x}_{t-1}, x_{t,i-1} \bigr]$ to compress the $i$-th control bit of the $t$-th \gls{tti}. A memory buffer of size $L$ is assumed and the output layer can be a single neuron with a sigmoid activation function $f_\text{sigmoid}(\hat{x}_{t,i}) = \frac{1}{1+e^{-\hat{x}_{t,i}}}$ that represents the probability of the $i$-th control bit being $1$. 
However, this method, particularly with \glspl{rnn}, faces scalability issues due to linearly increasing time indices with the \gls{tti} length, risking gradient vanishing and reduced performance. Moreover, the bit-wise compression approach necessitates $N$ invocations of the \gls{rnn} cell per \gls{dci} message, leading to significant computational overhead.
To mitigate these latency and scalability challenges, employing a transformer model is advantageous. 

Transformers efficiently handle long-range dependencies using positional encoding and attention mechanisms, making them well-suited for \gls{dci} compression, even with longer message lengths or larger memory buffers.
Moreover, without introducing much additional computational complexity, each field can be represented by an integer value and embedded by a neural embedding layer. The embedding layer maps each field into an arbitrary preset dimension to explore the correlations between fields.
In particular, depending on the hardware memory size of a \gls{bs} and a \gls{ue}, the integer embedding layer for the source binary data of a transformer network can be manually adjusted by a preset parameter, $\eta$. Then, the number of integer representations of one field can be calculated as
\begin{equation}
\vspace{-1mm}
	  s_{k} =
	\begin{cases}
		q_k2^{\eta} + 2^{\hat{\eta}} & \text{, } M_k > \eta, \\
		2^{M_k} & \text{, } M_k \leq \eta,
	\end{cases} 
\end{equation}  
for $k \in \{1,2,...,D\}$, where $q_k$ is the quotient of $M_k$ divided by $\eta$, and $\hat{\eta} = (M_k \text{ mod } \eta)$. Obviously, choosing $\eta = \max_{k \in \{1,2,...,D\}} \bigl{\{}M_k \bigr{\}}$ guarantees that all the \gls{dci} fields can be represented by a single integer and embedded. Otherwise, the \gls{dci} field is divided into several segments and transformed into an integer separately. Since the dictionary size for embedding may become large for a wireless system with a wide bandwidth which correspondingly needs control bits to represent the frequency domain allocation, choosing an affordable value of $\eta$ balances the tradeoff between the requirement of computational resources and compression performance. 
In total, there are $\sum_{k=1}^{D}s_k$ integers to be embedded by the embedding layer and the source binary data $\mathbf{x}_t$ can be represented by an integer form of $\tilde{\mathbf{x}}_t = \big{[}r_1, r_2, ..., r_R \big{]}$ with $R = \sum_{k=1}^D{(q_k + 1)}$, which refers to the number of integers required to encode the binary source data. 

With the integer representation of source \gls{dci} data, the encoder and decoder models of a transformer can be used to explore the temporal and spatial correlations, respectively. 
Denote $u^\text{encoder}_{t,k}$ and $u^\text{decoder}_{t,k}$ as the \gls{td} and \gls{sd} features for the encoder and decoder. 
A memory buffer of size $L$ can be used to form up the encoder feature.
The decoder feature can be constructed by memorizing all the previous bits that have been compressed and applying zero-padding to fix the size of feature.
As a result, $u^\text{encoder}_{t,k}$ and $u^\text{decoder}_{t,k}$ are constructed by:
\begin{align}
	\label{eq:encoder_feature}
	u^\text{encoder}_{t,k} &= \big{[} \tilde{\mathbf{x}}_{t-1}, \tilde{\mathbf{x}}_{t-2}, ...,\tilde{\mathbf{x}}_{t-L}  \big{]}, \\
	\label{eq:decoder_feature}
	u^\text{decoder}_{t,k} &= \big{[} r_1, r_2,...,r_{k-1},  \mathbf{0}_{\sum_{j=1}^{k-1}(q_j+1)} \big{]}.
 \vspace{-3mm}
\end{align}

\begin{table}[t!]
	\vspace{-3mm}
	\begin{algorithm}[H]
		\normalsize
		
		\caption{DCI Compression with a transformer model}
		\label{algo:gen_feature}
		\begin{algorithmic}
			\State \textbf{Input:} $\tilde{\mathbf{x}}_t$, $\tilde{\mathbf{x}}_{t-1}$,
			...,$\tilde{\mathbf{x}}_{t-L}$
			\State \textbf{Output:} $\dot{\mathbf{x}}_t$
			\State \textbf{Step 1: } Generate TD \& SD features: encoder and decoder features are found by (\ref{eq:encoder_feature}) and (\ref{eq:decoder_feature}), respectively.
			\State \textbf{Step 2: } Find the probability estimate for each bit in the $k$-th field: $\hat{\mathbf{y}}_{t,k} = f_\text{transformer}(u^\text{encoder}_{t,k}, u^\text{decoder}_{t,k})$.
			\State \textbf{Step 3: } Apply AC and compress the bits in the $k$-th field: $f_\text{AC}(\hat{y}_{t,k,j})$ for $j \in \{m | x_m \in \mathbf{d}_{t,k}\}$.
			\State \textbf{Step 4: } Move to the next field: $k \leftarrow k + 1$ and go back to Step 1.
		\end{algorithmic}
	\end{algorithm}
	\vspace{-7mm}
\end{table}

Regarding the output layer of a transformer model, conventional method employs a $\text{softmax}(\cdot)$ function to find the next possible symbol. 
For DCI compression, the objective function is to minimize the compression ratio, which equivalent to minimizing the cross-entropy loss between the estimate and the actual control bits.
Therefore, the output layer is designed to have a size of $S_\text{output} = \max_{k \in \{1,2,...,D\}}\bigl{\{}M_k \bigr{\}}$.
And the activation function can be adjusted to a $f_\text{sigmoid}(\cdot)$ function to estimate the distribution of each bit in the field.
Since the field size is a prior knowledge for both BS and \gls{ue}, during the sequential process of AC, the output neurons that are not valid for a field can be masked out to stabilize and improve the training performance. 
As a result, the output label for each \textbf{field} is defined as 
\begin{equation}
\mathbf{y}_{t, k} = \big{[} \mathbf{d}_{t,k}, \mathbf{0}_{S_\text{output} - M_k} \big{]},
\vspace{-1mm}
\end{equation} 
for $k \in \{1,2,...,D\}$.
In summary, the major block sizes for compressing a \gls{dci} message with a transformer model are proposed as:

\begin{equation}
\vspace{-1mm}
	\label{eq:layers}
	\begin{aligned}
		S_\text{encoder} &= LR \\
		S_\text{decoder} &= R \\
		S_\text{output} &= \max_{k \in \{1,2,...,D\}}\bigl{\{}M_k \bigr{\}}.
	\end{aligned}
 \vspace{-1mm}
\end{equation}

Representing the \gls{dci} message by fields and predicting the field value reduces the latency of compression and decompression since the NN models are called by much fewer times than bit-wise processing.

\graphicspath{{./resources/}}
\begin{figure}[t!]
	\centering\vspace{-3mm}
	\includegraphics[width=0.95\columnwidth]{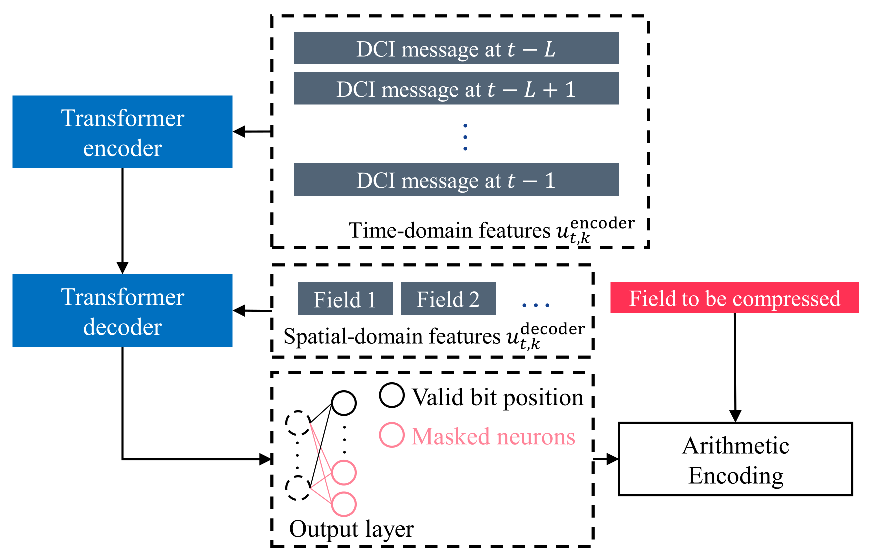}
	\caption{DCI compression with a transformer model}
	\label{fig:compression_flowchart}
\vspace{-6mm}
\end{figure}

\vspace{-7pt}
\subsection{DCI compression}
(\textbf{Training phase})
Given a database of DCI messages, the features and labels for the transformer model are firstly generated. 
To update the trainable parameters, a \gls{bce} loss can be computed by:
\vspace{-1mm}
\begin{equation}
    \begin{aligned}
	f_\text{BCE}(y_{t,k}, \hat{y}_{t,k}) = & \frac{1}{S_\text{output}} \sum^{S_\text{output}}_{j=1} y_{t,k,j}\text{log}(\hat{y}_{t,k,j}) \\
    & + (1-y_{t,k,j})\text{log}(1-\hat{y}_{t,k,j}).
    \end{aligned}
\end{equation}

A validation set is pre-defined from the training data to evaluate the performance of the trained model. The model with the lowest BCE loss of the validation set is saved and used in the test phase.
Note that the \gls{bts} can independently manage the training phase by storing transmitted \glspl{dci} in a buffer to create a training dataset. The well-trained model can be distributed to the \gls{ue} before initiating a session that utilizes \gls{dci} compression.

(\textbf{Test phase}) Once the transformer model has been well-trained, the trainable parameters are frozen in the test phase.
Let $f_\text{transformer}(\cdot) \in \mathbb{R}^{S_\text{output}}$ be the function of a transformer model, which takes $u^\text{encoder}_{t,k}$ and $u^\text{decoder}_{t,k}$ as the inputs and returns $\hat{\mathbf{y}}_{t,k}$ as the estimate of $\mathbf{y}_{t, k}$.
Following the structure of AC, each bit and the probability distribution $\hat{y}_{t,k,j}$ for $j \in \{m|x_m \in \mathbf{d}_{t,k}\}$ within the field is sequentially added to the AC encoder (decoder) to form the lossless compressed binary sequence. 
As shown in Fig. \ref{fig:compression_flowchart}, the transformer model is called for sequential compression of each field. 
The steps of achieving DCI lossless compression with a transformer model is summarized in Algorithm \ref{algo:gen_feature}, where $\dot{\mathbf{x}}_t \in \mathbb{F}^{K_t}_2$ denotes the compressed binary sequence for the $t$-th DCI message, with $K_t$ as the final length of a compressed sequence. 
After compression, a \gls{crc} is added to the \gls{dci} message and it is encoded for transmission on the \gls{pdcch}. 
The receiver uses the same transformer model to losslessly decompress the \gls{dci} message.

\vspace{-6pt}
\subsection{\gls{dci} field sorting}
\begin{figure}[t!]
	\centering\vspace{-3mm}	\includegraphics[width=0.95\columnwidth]{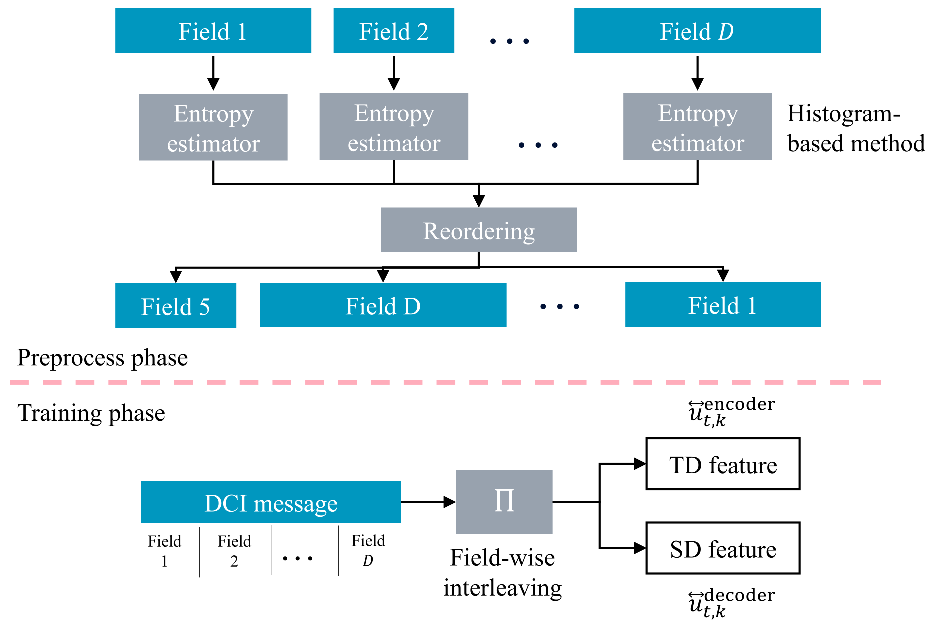}
 \vspace{-3mm}
	\caption{Reordering DCI fields to reach a better convergence result}
	\label{fig:reordering}
	\vspace{-5mm}
\end{figure}

A common practical \gls{ml} problem is choosing the right model that fits well to the feature data, or adjusting the feature data without introducing much computational complexity so that the chosen model can learn the features faster and converges to a good generalization performance.
For ML-based lossless compression, the compression performance relies on how well the learned model matches to the true underlining distribution of each symbol to be compressed conditioned on the given features. 
As can be noticed from (\ref{eq:decoder_feature}), the features of a transformer's decoder follow a structure of Toeplitz matrix, where the left-most fields are used as features for more times than the tail field. 
This comes from the fact that the \gls{ac} follows a sequential encoding (decoding) manner \cite{arithmetic_coding}.
The first encoded (decoded) fields are treated as the prior knowledge for the following fields. 
In other words, the distribution of features for the transformer model to learn from depends on the order of \gls{dci} fields.
Therefore, we propose to reorder the fields on top of the proposed \gls{dci} compression scheme. 
If the number of \gls{dci} fields is small, a solution to determine the field order is listing out all the possible combinations and train the transformer model to find out the best model.
However, listing out all the possible combinations is infeasible as the number of combinations is $D!\ $ and the number of \gls{dci} fields is generally more than 10.
To determine an order without introducing much additional complexity, we propose to order the fields with a descending order of entropy value of the field.
Denote $\mathcal{E}_k$ as the alphabet of set of discrete values and $E_k$ as the underlining variable for the $k$-th \gls{dci} field in the training dataset. 
Then, the entropy value can be estimated from a histogram-based method that has
\begin{equation}
 \vspace{-3mm}
	\label{eq:entropy}
	H(E_k) = -\sum_{e\in \mathcal{E}_k}^{}p(e)\text{log}\Big{(}p(e)\Big{)},
\end{equation}
where $p(e)$ denotes the normalized frequency of discrete value $e$.
Since the field comprises binary bits, the entropy of each field can be estimated through a histogram-based method as in (\ref{eq:entropy}).
Once the entropy of each field is estimated, the fields are sorted in a descending order, where the more uncertainty (randomness) of the field, the closer to the front that the field is allocated to.
A direct intuition behind it is that the field with a large entropy may contain more bits. 
The \gls{sd} features follow a Toeplitz matrix, where the more randomness of the field, the more frequent of the field that will be used in the transformer's decoder feature $\hat{u}_{t,k}^\text{decoder}$.
Diversified features avoid bias information that might lead the transformer model to be trained to a local minimal.
In contrast, if the control bits are constant, which result in a low entropy value, putting these bits to the front may easily guide the transformer to converge to a non-generalized model due to the bias of bit value. The proposed field-wise interleaving method is considered as a preprocessing step before training the transformer model as shown in Fig. \ref{fig:reordering}.
With the preprocessed order that is computed by (\ref{eq:entropy}) from the training dataset for each field, \gls{dci} fields are interleaved and all the features and labels collection and the \gls{dci} compression steps can follow Algorithm \ref{algo:gen_feature}.

\graphicspath{{./resources/}}

\vspace{-3pt}
\section{Numerical Results}
\vspace{-1mm}
In this section, we provide the simulation results for \gls{dci} compression with a comparison among the algorithms of \gls{hc}, \gls{rnn}-based DeepZip and our proposed lossless compression technique with a transformer model.
Besides showing the compression ratios as the performance metric, we simulated \gls{pdcch} encoding and decoding with an assumption of \gls{awgn} channel and verified that \gls{dci} compression is a powerful technique to improve the reliability of \gls{pdcch}.

To create the database of \gls{dci} messages, we first use a Matlab system level simulator to generate a scheduling log. 
Based on the scheduling log, corresponding field values in a \gls{dci} message is assigned. 
The key system parameters are summarized in Table \ref{table:system}, where $\sim \mathcal{U}(10, 30)$ denotes that the data rate of each \gls{ue} is randomly sampled from a uniform distribution within the interval of 10 to 30 Mbps. The transformer model follows the conventional architecure in \cite{transformer}, where 4 multi-head attentions are used and there are 64 neurons in the embedding layer. Adam optimizer is utilized to update the trainable parameters.

\vspace{-8pt}
\subsection{Compression ratios}
\begin{figure}[t!]
	\graphicspath{{./resources/numerical_results/}}
	\begin{subfigure}{0.45\columnwidth}
		\centering
		\includegraphics[width=\textwidth]{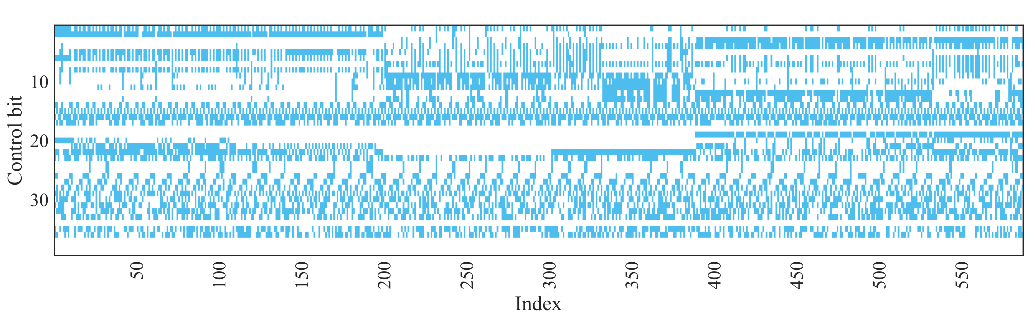}
		\caption{Original DCI}

		\label{fig:original_DCI}
	\end{subfigure}\hfill
	\begin{subfigure}{0.45\columnwidth}
		\centering
		\includegraphics[width=\textwidth]{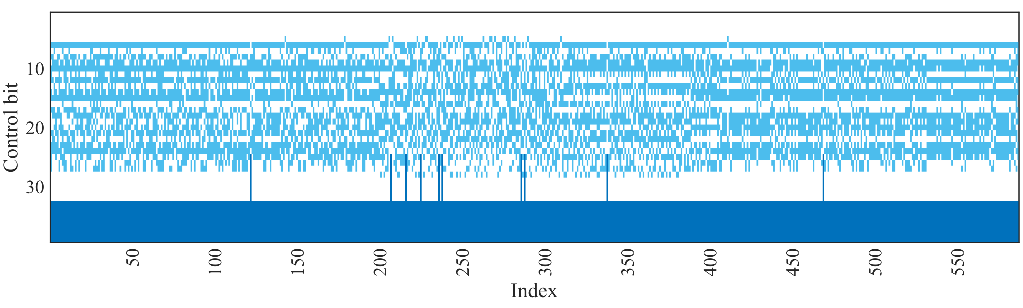}
		\caption{Huffman Coding}
		\label{fig:HC}
	\end{subfigure}

	\begin{subfigure}{0.45\columnwidth}
		\centering
		\includegraphics[width=\textwidth]{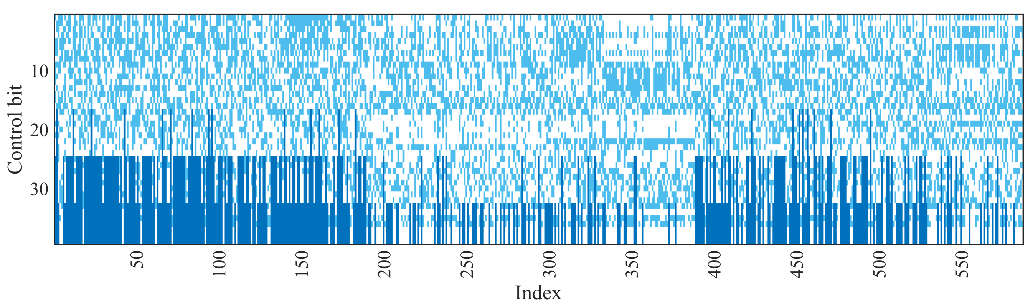}
		\caption{\gls{rnn}-DeepZip}
		\label{fig:RNN}
	\end{subfigure}\hfill
	\begin{subfigure}{0.45\columnwidth}
		\centering
		\includegraphics[width=\textwidth]{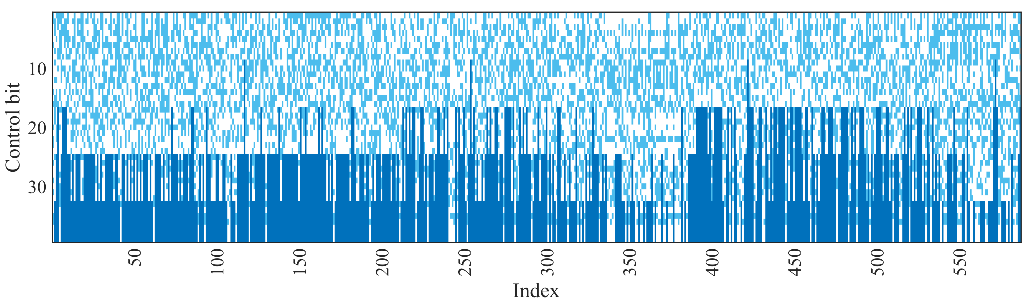}
		\caption{Transformer-based}
		\label{fig:transformer}
	\end{subfigure}
	
	\begin{subfigure}{0.45\columnwidth}
		\centering
		\includegraphics[width=\textwidth]{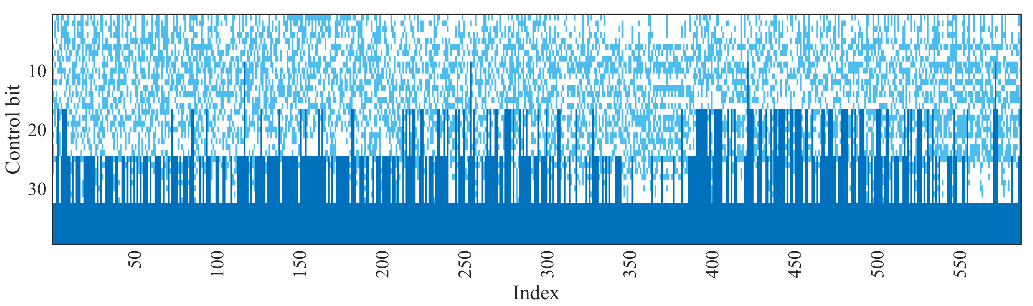}
		\caption{Joint Transf. \& HC}
		\label{fig:transformer_hc}
	\end{subfigure}
	
	\caption{Comparison on compression ratios, where the light blue dot indices a control bit with bit value of 1 and the white dot refers to 0. The dark blue dot refers to the null space.}
	\label{fig:compression_comparison}
\vspace{-4.5mm}
\end{figure}

We simulated a network, wherein 3 \glspl{ue} are scheduled per \gls{tti}. We also asume 13 available resource block groups in the downlink.
We then collected along trace of \gls{dci} messages for each \gls{ue}, and applied a (97\%, 3\%) split, where the last 3\% of all \gls{dci} messages for each \gls{ue} are used for testing purposes.
``HC'' and ``RNN-DeepZip'' refer to the \gls{hc} and RNN-based lossless compression methods.
``Transformer-based'' refer to the proposed transformer model for \gls{dci} compression. 
Note that there is another method listed in Fig. \ref{fig:transformer_hc}, named ``Transformer \& \gls{hc} '', which combines the transformer-based lossless compression and \gls{hc}. 
Since \gls{hc} is observed to provide a stable lossless compression performance,
when the Transformer-based method does not achieve a shorter \gls{dci} length, \gls{hc} is performed. 
During the training phase, 3 models are trained separately for each \gls{ue}. 
And in the test phase, the trainable parameters are frozen.

\begin{figure}[t!]
	\centering	\includegraphics[width=0.75\columnwidth]{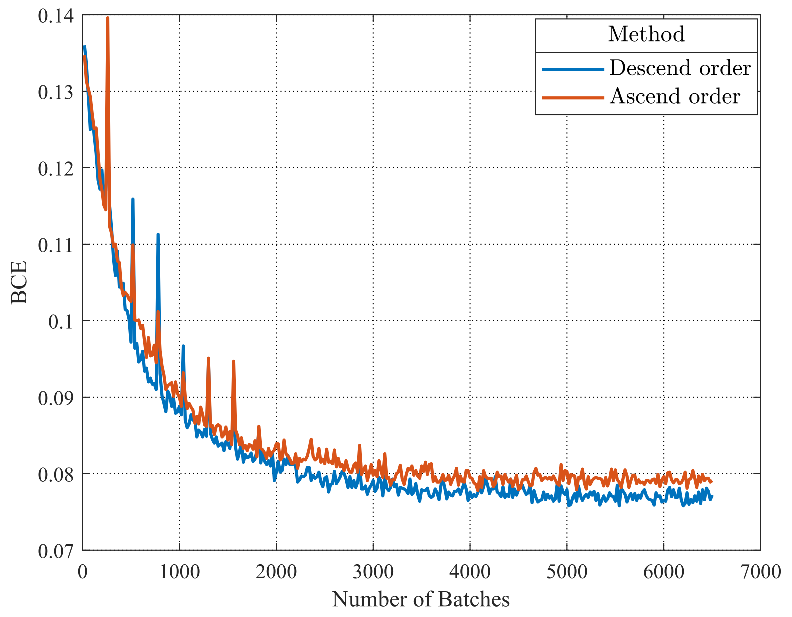}\vspace{-3mm}
	\caption{Training curves comparison by ordering the fields' entropies by a descending order and an ascending order}
	\label{fig:training_curve}
	\vspace{-4mm}
\end{figure}

The \gls{dci} message has a payload length of 39. An average compression ratio is used as the performance metric to describe the compression performance. 
The average compression ratio is defined as the original \gls{dci} length over the compressed \gls{dci} length. 
To have a broad view of compression performance over all the \glspl{ue}, Fig. \ref{fig:compression_comparison} concatenates the test \gls{dci} messages for all the 3 \glspl{ue}. As shown in Fig. \ref{fig:compression_comparison} and Table \ref{table:comp_ratio}, RNN-DeepZip, Transformer-based and Joint Transf. \& \gls{hc} all outperform HC on the average compression ratio. 
It can be observed that a transformer-based \gls{dci} compression technique achieves a better compression ratio than \gls{hc} and \gls{rnn}-DeepZip. In particular, ``Joint Transf \& \gls{hc}'' has a compression ratio around $28.3\%$ higher than \gls{hc}. Additionally, Fig. \ref{fig:training_curve} demonstrates the effectiveness of reordering the fields by a descend order of entropies compared to an ascending order.

\begin{table}[t!]
	\centering
        \caption{System parameters for generating scheduling logs}
        \vspace{-2mm}
	\begin{center}
		\begin{tabular}{c| c }
			\hline
			Parameters & Value \\
			\hline\hline
			Number of UEs &  3  \\ \hline
			Number of RBGs   & 13 \\ \hline
			Scheduler   & Proportional Fair \\ \hline
			Traffic model & On-Off network traffic \\
			\hline
			Application data rate & $\sim \mathcal{U}(10, 30)$ Mbps \\
			\hline
		\end{tabular}
	\end{center}
	\label{table:system}
	\vspace{-2mm}
\end{table}

\begin{table}[t!]
	\centering
        \vspace{-2mm}
        \caption{Average compression ratio}
        \vspace{-3mm}
	\begin{center}
		\begin{tabular}{c| c }
			\hline
			Methods & Avg. compression ratio \\
			\hline\hline
			\gls{hc} &  1.2   \\ \hline
			\gls{rnn}-DeepZip   & 1.23 \\ \hline
			Transformer-based  & 1.46 \\ \hline
			Joint Transf. \& \gls{hc}  & 1.54 \\
			\hline
		\end{tabular}
	\end{center}
	\vspace{-1mm}
	\label{table:comp_ratio}
	\vspace{-2mm}
\end{table}

\vspace{-2mm}
\subsection{Channel decoding performance}
To demonstrate the potential decoding performance gain with lossless compression, we simulated polar-encoded \gls{pdcch} with an encoded length of 128. Zero-paddings are appended to the payload after lossless compression. 
Given the histogram of compression length, we randomly sample a payload length and add zeros to the end of compressed data to form-up the final payload before channel encoding. 
With the histogram of compressed length by HC and a list decoding algorithm that lists out the possible number of zero-paddings to the DCI payload, the decoding performance can be improved by 0.65 dB at an FER of $10^{-2}$.
When ``Joint Transf. \& \gls{hc}'' is performed, a total of 0.8 dB gain can be obtained.

\graphicspath{{./resources/numerical_results/}}

\graphicspath{{./resources/numerical_results/}}
\begin{figure}[t!]
\centering\includegraphics[width=0.75\columnwidth]{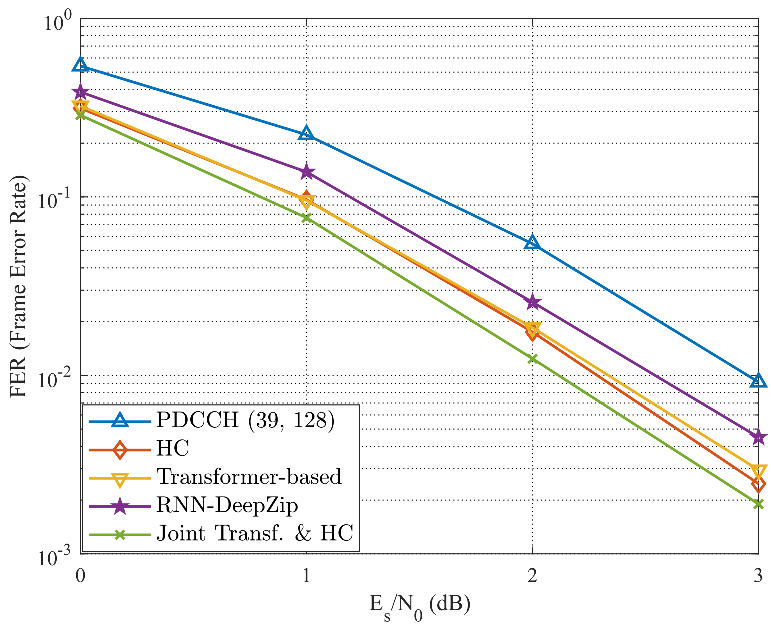}\vspace{-3mm}
	\caption{Frame error rate comparison with lossless compression over an \gls{awgn} channel}
	\label{fig:fer}
	\vspace{-4.5mm}
\end{figure}

\vspace{-2pt}
\section{Conclusion}
This paper proposes a transformer-based lossless compression technique for \gls{dci} messages. 
Besides proposing a framework to losslessly compress the length of a \gls{dci} message, a sorting mechanism for \gls{dci} fields is proposed to further improve the compression ratio.
The proposed architecture explores both spatial and temporal correlations. 
With the reduced \gls{dci} length and code rate, a more reliable control channel can be achieved and potentially an improved channel capacity can be obtained by controlling more \glspl{ue} in a limited time.

\vspace{-1mm}
\section*{Acknowledgements}
\vspace{-1mm}
The work is funded by the European Union through project CENTRIC (G.A no. 101096379).

\end{document}

%% file: acronyms.tex
\newacronym{bts}{BTS}{Base Transceiver Station}
\newacronym{dci}{DCI}{Downlink Control Information}
\newacronym{crc}{CRC}{Cyclic Redundancy Check}
\newacronym{ml}{ML}{machine learning}
\newacronym{ac}{AC}{Arithmetic Coding}
\newacronym{hc}{HC}{Huffman Coding}
\newacronym{rnn}{RNN}{Recurrent Neural Network}
\newacronym{tti}{TTI}{Transmission Time Interval}
\newacronym{ue}{UE}{user equipment}
\newacronym{nn}{NN}{neural network}
\newacronym{bs}{BS}{base station}
\newacronym{pdcch}{PDCCH}{Physical Downlink Control Channel}
\newacronym{awgn}{AWGN}{Additive White Gaussian Noise}
\newacronym{bce}{BCE}{binary cross-entropy}
\newacronym{td}{TD}{time domain}
\newacronym{sd}{SD}{spatial domain}